\documentclass[manuscript]{aastex61}

\shorttitle{Filtering solar-like oscillations}
\shortauthors{Chaplin et al.}

\begin{document}

\title{Filtering solar-like oscillations for exoplanet detection in radial velocity observations}

\correspondingauthor{W. J. Chaplin}
\email{w.j.chaplin@bham.ac.uk}

\author{William J. Chaplin}
\affiliation{School of Physics and Astronomy, University of Birmingham, Edgbaston, Birmingham B15 2TT, UK}
\affiliation{Stellar Astrophysics Centre, Department of Physics and Astronomy, Aarhus University, Ny Munkegade 120, 8000 Aarhus C, Denmark}

\author{Heather M. Cegla}
\affiliation{Observatoire de Gen\`eve, Universit\'e de Gen\`eve, 51 Chemin des Maillettes, 1290, Versoix, Switzerland}
\affiliation{CHEOPS Fellow, SNSF NCCR-PlanetS}

\author{Christopher A. Watson}
\affiliation{Astrophysics Research Centre, School of Mathematics and Physics, Queen's University Belfast, University Road, Belfast, BT7 1NN, UK}

\author{Guy R. Davies}
\affiliation{School of Physics and Astronomy, University of Birmingham, Edgbaston, Birmingham B15 2TT, UK}
\affiliation{Stellar Astrophysics Centre, Department of Physics and Astronomy, Aarhus University, Ny Munkegade 120, 8000 Aarhus C, Denmark}

\author{Warrick H. Ball}
\affiliation{School of Physics and Astronomy, University of Birmingham, Edgbaston, Birmingham B15 2TT, UK}
\affiliation{Stellar Astrophysics Centre, Department of Physics and Astronomy, Aarhus University, Ny Munkegade 120, 8000 Aarhus C, Denmark}

\begin{abstract}

Cool main-sequence, sub-giant and red-giant stars all show solar-like oscillations, pulsations that are excited and intrinsically damped by near-surface convection. Many overtones are typically excited to observable amplitudes, giving a rich spectrum of detectable modes. These modes provide a wealth of information on fundamental stellar properties. However, the radial velocity shifts induced by these oscillations can also be problematic when searching for low-mass, long-period planets; this is because their amplitudes are large enough to completely mask such minute planetary signals. Here we show how fine-tuning exposure times to the stellar parameters can help efficiently average out the solar-like-oscillation-induced shifts. To reduce the oscillation signal to the radial velocity precision commensurate with an Earth-analogue, we find that for cool, low-mass stars (near spectral type K) the necessary exposure times may be as short as $\sim$4\,minutes, while for hotter, higher-mass stars (near spectral type F, or slightly evolved) the required exposure times can be longer than 100\,minutes. We provide guideline exposure durations required to suppress the total observed amplitude due to oscillations to a level of $0.1\,\rm m\,s^{-1}$, and a level corresponding to the Earth-analogue reflex amplitude for the star. Owing to the intrinsic stochastic variability of the oscillations, we recommend in practice choosing short exposure durations at the telescope and then averaging over those exposures later, as guided by our predictions. To summarize, as we enter an era of $0.1\,\rm m\,s^{-1}$ instrumental precision, it is critical to tailor our observing strategies to the stellar properties. 

\end{abstract}

\keywords{planets and satellites: detection --- stars: oscillations --- stars: low-mass --- Sun: granulation --- techniques: radial velocities --- methods: data analysis}

\section{Introduction} \label{sec:intro}


Since the dawn of the exoplanet field, one of the ultimate goals has
been to discover, confirm, and characterize a true Earth analogue;
thanks to advances in instrumental precision this will soon be
technically feasible. For example, the recent NASA \textit{Kepler}
mission has shown it is now possible to routinely detect Earth-sized
planet candidates. The \textit{Kepler} mission alone detected over
1000 candidates with a radius less than twice the Earth's, $R_p \le
2R_{\Earth}$, 379 of which had $R_p \le 1.25 R_{\Earth}$ and $\sim$20
of those are believed to be
temperate\footnote{exoplanetarchive.ipac.caltech.edu}. However,
without the masses for many of these candidates it is difficult to
confirm and further characterize their planetary nature. Fortunately,
the next generation of spectrographs currently being commissioned,
such as ESPRESSO on the VLT \citep{gonzalez17}, promise a radial
velocity (RV) precision\footnote{For first
  light details see https://www.eso.org/public/news/eso1739/ and
  https://news.yale.edu/2018/03/13/yales-expres-instrument-ready-find-next-earth-analog
  for ESPRESSO and EXPRESS, respectively.} of $0.1\,\rm m\,s^{-1}$, technically enabling the
future detection of an Earth-twin RV signal at $0.09\,\rm m\,s^{-1}$.

However, this increase in instrumental precision necessitates an increased understanding and treatment of astrophysical effects originating from the host stars. This is particularly troublesome for RV follow-up, because inhomogeneities on the stellar surface can alter the observed stellar absorption lines, thereby changing the line profiles' centre-of-light, which in turn can be mistaken for wholesale Doppler shifts that may mask or even mimic planetary signals \citep{saar97}. Most stellar surface phenomena manifest themselves in RV measurements through three main routes caused by: flux imbalances; convection and its (magnetic) suppression; and wholesale stellar surface shifts. Dark starspots emit less flux and therefore appear as `emission' bumps in absorption lines, and vice versa for bright faculae/plage, which ultimately alters the shapes of the observed line profiles \citep{vogt87, saar97}. Convection results in hot, bright granules rising to the surface, cooling, and falling back down into the inter-granular lanes; since the granules are larger and brighter, this results in a net convective blueshift and an asymmetric line profile that changes as the granules evolve (for the Sun, this blueshift is near 350 m~s$^{-1}$ at disc center and most lines have `C'-shaped bisectors; \citealt{gray08}). In regions of enhanced magnetic activity, the magnetic field lines can suppress the convection and reduce the net blueshift (in addition to altering the line shapes; \citealt{saar97, meunier10, dumusque14, haywood16}). Acoustic waves excited by convection in the near-surface layers of cool stars can set up internal standing waves. The resulting resonant p-modes (p indicating that gradients of pressure provide the restoring force) may be detected via the gentle oscillations they give rise to \citep{chaplin13}. For Sun-like stars the dominant timescales for these oscillations are of the order of minutes. Additionally, if the stellar radius changes (e.g. due to changes in convection over a magnetic activity cycle or in localized regions such as the Wilson depression in starspots), but the mass is conserved, then the gravitational redshift of the stellar lines, and therefore net RV shift, also changes \citep{cegla12}. 

Crucial to understanding the impact of such `astrophysical noise' sources on exoplanet detection are the amplitude and timescale of each of these stellar phenomena. On magnetically active stars, starspots and faculae/plage can typically induce RV shifts of 1-100~m~s$^{-1}$ \citep{saar97} and the suppression of convection can induce shifts of tens of m~s$^{-1}$ \citep{meunier10}. All stars with a convective envelope, both active and magnetically `quiet', exhibit RV shifts from p-modes and granulation at the level of tens of cm~s$^{-1}$ to m~s$^{-1}$, while variable gravitational redshift is likely to be on the $0.1\,\rm m\,s^{-1}$ level or lower \citep{cegla12}. Although there are numerous ongoing efforts to model and remove the contamination from larger amplitude effects like starspots, faculae/plage, and suppression of convection \citep[for a non-exhaustive list see][and references therein]{aigrain12, boisse09, boisse11, dumusque14, hatzes10, haywood14, haywood16, hebrard16, herrero16, lopezmorales16, meunier17}, the best candidates for temperate, rocky world detections are still around magnetically inactive stars. However, these stars will still show detectable astrophysical noise signatures due to granulation and p-modes at the m~s$^{-1}$ level (e.g., see \citealt{yu18, medina18}). To circumvent these noise sources, the most commonly used strategy to date is to bin-down the noise with an optimized observing strategy \citep{dumusque11a}. In particular, \cite{dumusque11a} argue for three measurements per night, separated by two hours, with ten-minute exposures; the purpose of multiple measurements in a given night is to tackle the supergranulation timescales and the purpose of the ten-minute exposures is to suppress contributions from the p-modes. As such, the same exposure length is generally used to bin-down the contribution from p-modes across a range of spectral types with varying masses, surface gravities, and effective temperatures (discounting exposure length variations due to system brightness
considerations). 

However, once spectrographs enable $0.1\,\rm m\,s^{-1}$ precision it will be desirable to tailor the observing strategy to suppress the p-mode signatures in particular on a star-by-star basis. This is because key parameters associated with the oscillations -- notably the timescale on which the most prominent modes are observed, and the oscillation amplitudes -- scale with fundamental stellar properties, and the signal can be filtered by an appropriate choice for the length of the exposures. This is in contrast to, for example, the signal given by granulation, which rises in amplitude at progressively lower frequencies (longer timescales) meaning a simple low-pass filter -- as given by lengthening exposure times -- is insufficient. It is worth adding that the total amplitude of the granulation signal is significantly lower than the total amplitude due to solar-like oscillations when both phenomena are observed in Doppler velocity.

Fine-tuning the exposure durations to most effectively average out the p-mode oscillations is the subject of this paper. We investigate the impact of various exposure lengths on data for cool stars of different spectral types, and make recommendations for the appropriate exposure lengths to use, depending on the stellar properties of the target.

The layout of the rest of the paper is as follows. In Section~\ref{sec:filter}, we introduce and discuss the characteristics of the frequency-domain filtering given by simple exposures of finite duration. Then in Section~\ref{sec:sun}, we use the solar p-mode oscillations spectrum as a test case to illustrate how changing the exposure duration affects the measured residual (filtered) amplitude of the oscillations. Section~\ref{sec:other} presents detailed results on optimal exposure durations for cool stars across the Hertzsprung-Russell diagram that are expected to show solar-like oscillations, including low-mass main-sequence, sub-giant and low-luminosity red-giant stars. To enable the community to use our results, we have made publicly available on \texttt{Github} a Python code to calculate optimal exposure durations given basic stellar observables as input. This code is described in Section~\ref{sec:how}. We conclude the paper in Section~\ref{sec:conc} with summary remarks.

%
%
%

 \section{Filter response in frequency}
 \label{sec:filter}
 
Empirical data have finite exposure times, which naturally create a boxcar filtering of the underlying astrophysical signatures. The classic boxcar filter is advantageous as it has a well-determined transfer function in the frequency domain (see, e.g., \citealt{chaplin14} and references therein). This means we can calculate the residual signal amplitudes that will remain after adopting various integration times, provided the amplitude-frequency content of the spectrum of p-mode oscillations is known.


As such, we explore integration times made at intervals $\Delta t$. Sampling theory tells us that the highest frequency (equivalent to the longest period) that can then be measured unambiguously is the Nyquist frequency\footnote{If the sampling of the data in the time domain is irregular, the median sampling provides a good estimate of the notional Nyquist frequency.}, which is defined as
 \begin{equation}
 \nu_{\rm Nyq} \stackrel{\mathrm{def}}{=} (2\Delta t)^{-1}.
 \label{eq:nyq}
 \end{equation}
If the integration time per cadence is $\Delta t_{\rm c}$, i.e., the amount of time during each cadence $\Delta t$ that data are collected (so that $\Delta t_{\rm c} \le \Delta t$), then a signal of frequency $\nu$ will have its amplitude attenuated by the factor
 \begin{equation}
 \eta(\nu) = {\rm sinc} \left[ \pi \left( \nu \Delta t_{\rm c} \right) \right],
 \label{eq:sincamp}
 \end{equation}
while the response in \textsl{power} will be attenuated by:
 \begin{equation}
 \eta^2(\nu) = {\rm sinc}^2 \left[ \pi \left( \nu \Delta t_{\rm c} \right)
   \right].
 \label{eq:sincpow}
 \end{equation}
In what follows we shall assume that data quality is comparable from one integration to the next. 
 
\begin{figure*}
 \plotone{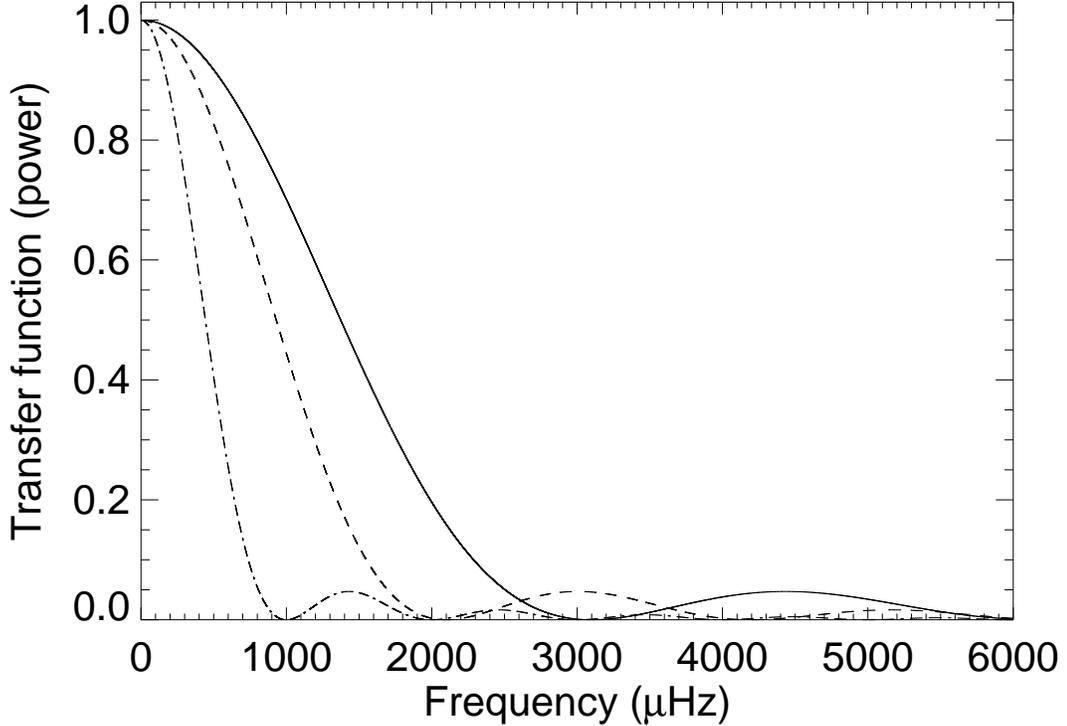}

 \caption{\small The transfer functions (in power) given by exposure durations $\Delta t_{\rm c}$ of $5.4\,\rm min$ (solid line), 7.9\,min (dashed line) and 16.7\,min (dot-dashed line).}

 \label{fig:filt1}
\end{figure*}

The transfer functions in power (i.e. Equation~\ref{eq:sincpow}) given by exposure durations $\Delta t_{\rm c}$ of $5.4\,\rm min$ (solid line), 7.9\,min (dashed line) and 16.7\,min (dot-dashed line) are shown in Fig.~\ref{fig:filt1}. The reason for selecting these durations will become apparent below in Section~\ref{sec:sun}.



\newpage

\section{Results for the Sun or a solar twin}
\label{sec:sun}

\begin{figure}
 \plottwo{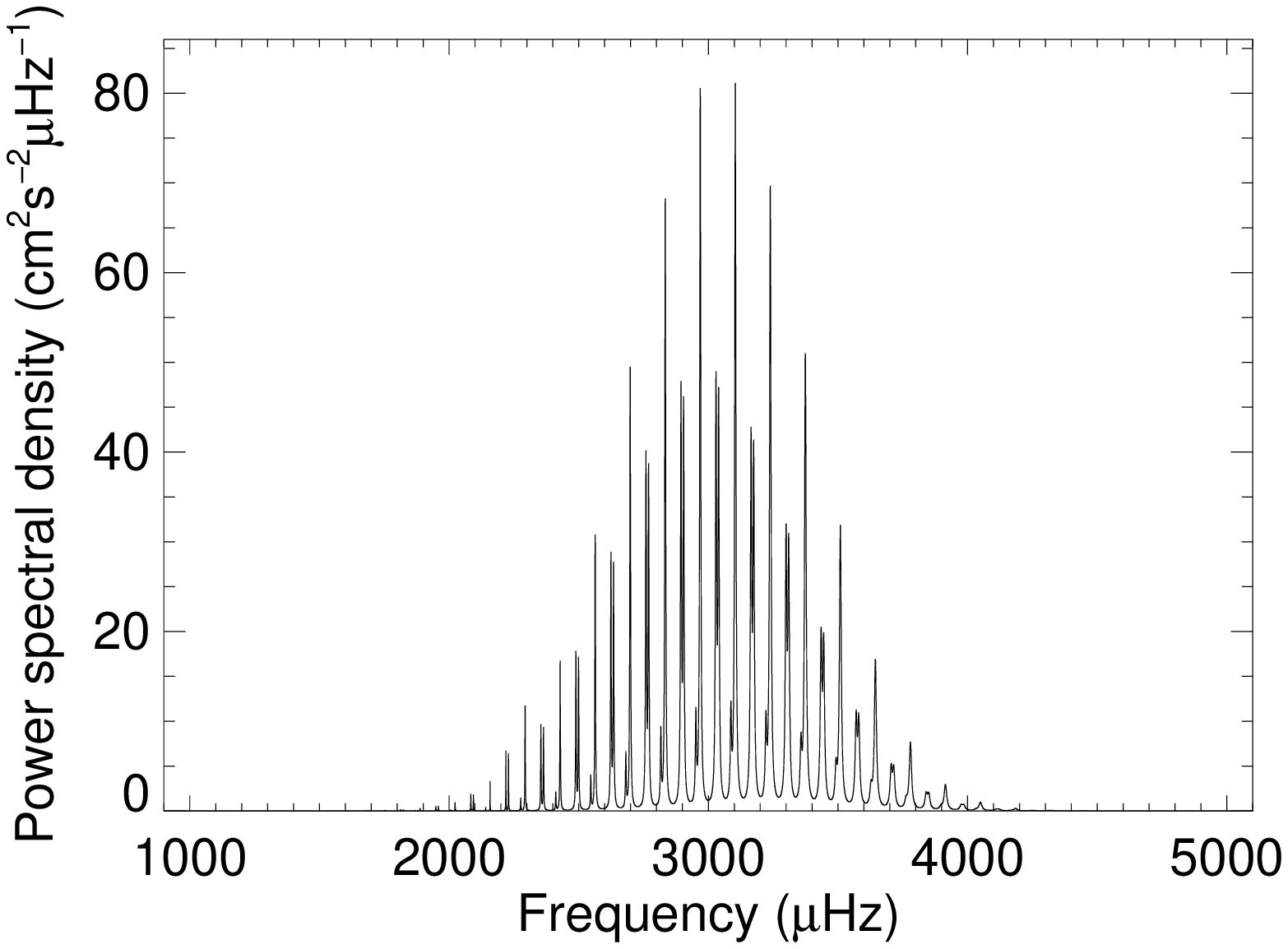}{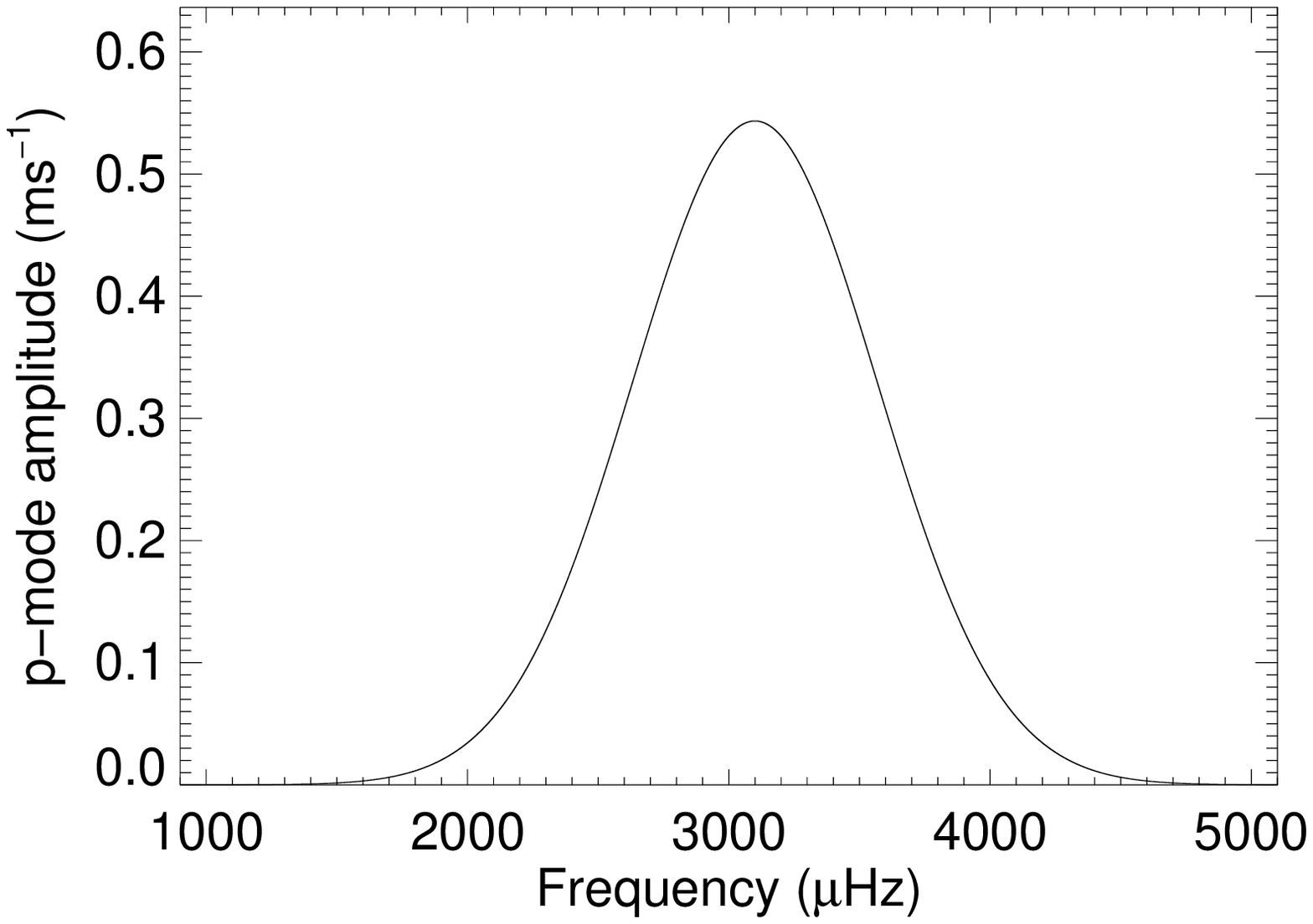}
 \caption{\small Left-hand panel: Model oscillation spectrum, constructed to mimic the spectrum shown by Sun-as-a-star observations. Right-hand panel: mode amplitude, $\sqrt{P_n}(\nu)$ (Equation~\ref{eq:pnnu}), as a function of frequency. Note both panels have been
calibrated to show the full power spectral density and amplitude, respectively, as opposed to the mean-square and root-mean-square values.}

 \label{fig:solar1}
\end{figure}

To test the filter response for a Sun-like star, we first constructed a model p-mode oscillation spectrum to mimic Sun-as-a-star observations; this is shown in the left-hand panel of Fig.~\ref{fig:solar1}. This model spectrum comprises many overtones, $n$, of modes of different angular (spherical) degree, $l$. The solar p modes are stochastically excited and intrinsically damped by turbulence in the near-surface layers of the convective envelope, and manifest themselves as Lorentzian-like peaks in the frequency spectrum. Other stars having outer convective envelopes show similar oscillation spectra, where many overtones can be excited to detectable amplitudes.

Modes of different degree $l$ show different powers at the same frequency owing to the net averaging over the visible stellar hemisphere of perturbations due to the different spherical harmonics. The relative visibilities also depend on details of the method used to make the observations, since this can affect the spatial sensitivity weighting of the over the stellar disc (e.g., see \citealt{jcd89, basu17}; and further comments below).

Within any given order, $n$, the total observed mode power is given by:
  \begin{equation}
  P_n = A^2_{n0} \sum_{l} \left( S_l/S_0 \right)^2,
  \label{eq:pn}
  \end{equation}
where $A_{n0}$ is the full amplitude\footnote{Here, we deal with full, not RMS, amplitudes.} of the radial ($l=0$) mode in that order, and the sum is over the relative visibilities of the modes of different degree, $l$, that appear in the order (those visibilities $S_l$ being normalized with respect to the radial-mode visibility $S_0$, all expressed in amplitude).  Signatures of modes of $0 \le l \le 3$ are readily detectable in Sun-as-a-star (i.e., non-disc-resolved) observations. Whilst weaker signatures due to modes of $l=4$ and even $l=5$ are also just discernible, their relative contribution to the observed $P_n$ is negligible.

Both $P_n$ and $\sqrt{P_n}$ -- the latter being the total p-mode amplitude in each order -- follow, to good approximation, a Gaussian in frequency (as shown in Fig.~\ref{fig:solar1}). The resulting envelope of power or amplitude is centered on a frequency we call $\nu_{\rm max}$. If $A_{\rm max}$ is the equivalent maximum radial-mode amplitude at this central frequency, we may describe the variation with frequency of $P_n$ according to:
 \begin{equation}
 P_n(\nu) = A_{\rm max}^2 \sum_{l} \left( S_l/S_0 \right)^2
            \exp\left(- \frac{(\nu-\nu_{\rm max})^2}{2c_{\rm env}^2} \right),
 \label{eq:pnnu}
 \end{equation}
where $c_{\rm env}$ fixes the width of the Gaussian envelope. Since $P_n(\nu)$ is assessed an order at a time -- i.e., at separations in frequency corresponding to the overtone spacing between the modes, the so-called large frequency separation, $\Delta\nu$ -- it is a smooth function in frequency. The amplitude $\sqrt{P_n}(\nu)$ is plotted in the right-hand panel of Fig.~\ref{fig:solar1}.

Several long-standing observational programmes have been dedicated to collecting data in Doppler velocity for helioseismology studies. Examples providing Sun-as-a-star data are the ground-based Birmingham Solar-Oscillations Network (BiSON; \citealt{chaplin96, hale16}) and the Global Oscillations at Low Frequency instrument (GOLF; \citealt{gabriel95}) on board the ESA/NASA SoHo spacecraft \citep{domingo95}. Both instruments make their observations by measuring the relative intensities in narrow passbands in the blue and red wavelength wings of a single spectral line. This is in marked contrast to stellar spectrographs, which use many lines and cross-correlate the observed line profiles with standard reference or synthetic spectra. These crucial differences affect the observed $A_{\rm max}$ and $S_l$. Because our predictions for other stars (see Section~\ref{sec:other}) are calibrated against Sun-as-a-star observations, we have deliberately chosen to use solar radial-mode amplitudes and relative visibilities in our model that are consistent with those expected for observations made using a spectrograph like HARPS (see\footnote{Also \citet{palle13} for solar observations made by the SONG \citep{grundahl08} Hertzsprung telescope.} Table~1 in\ \citealt{kjeldsen08}), which differ slightly from the BiSON and GOLF values (see \citealt{basu17} for further discussion).


\begin{figure}
 \plottwo{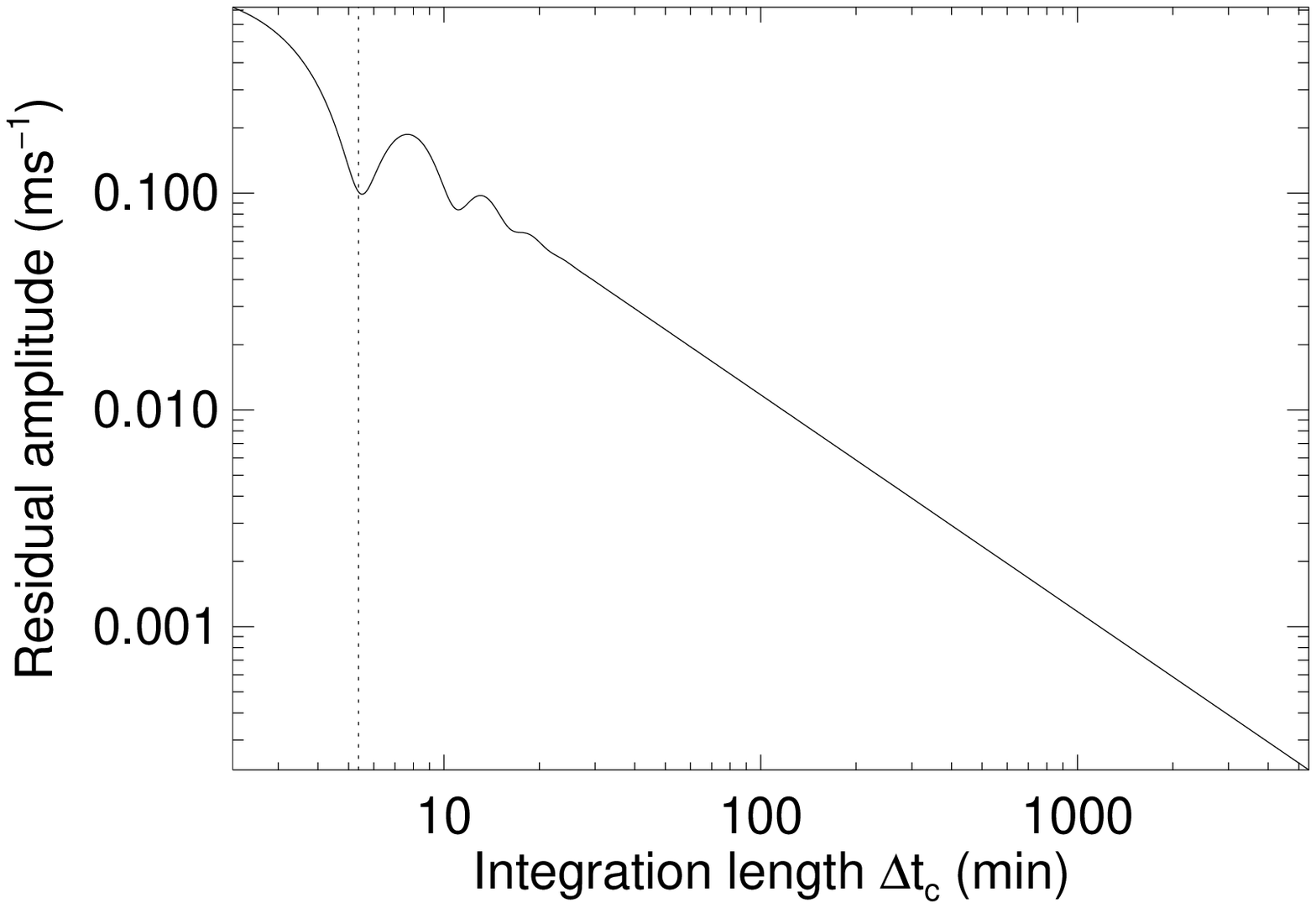}{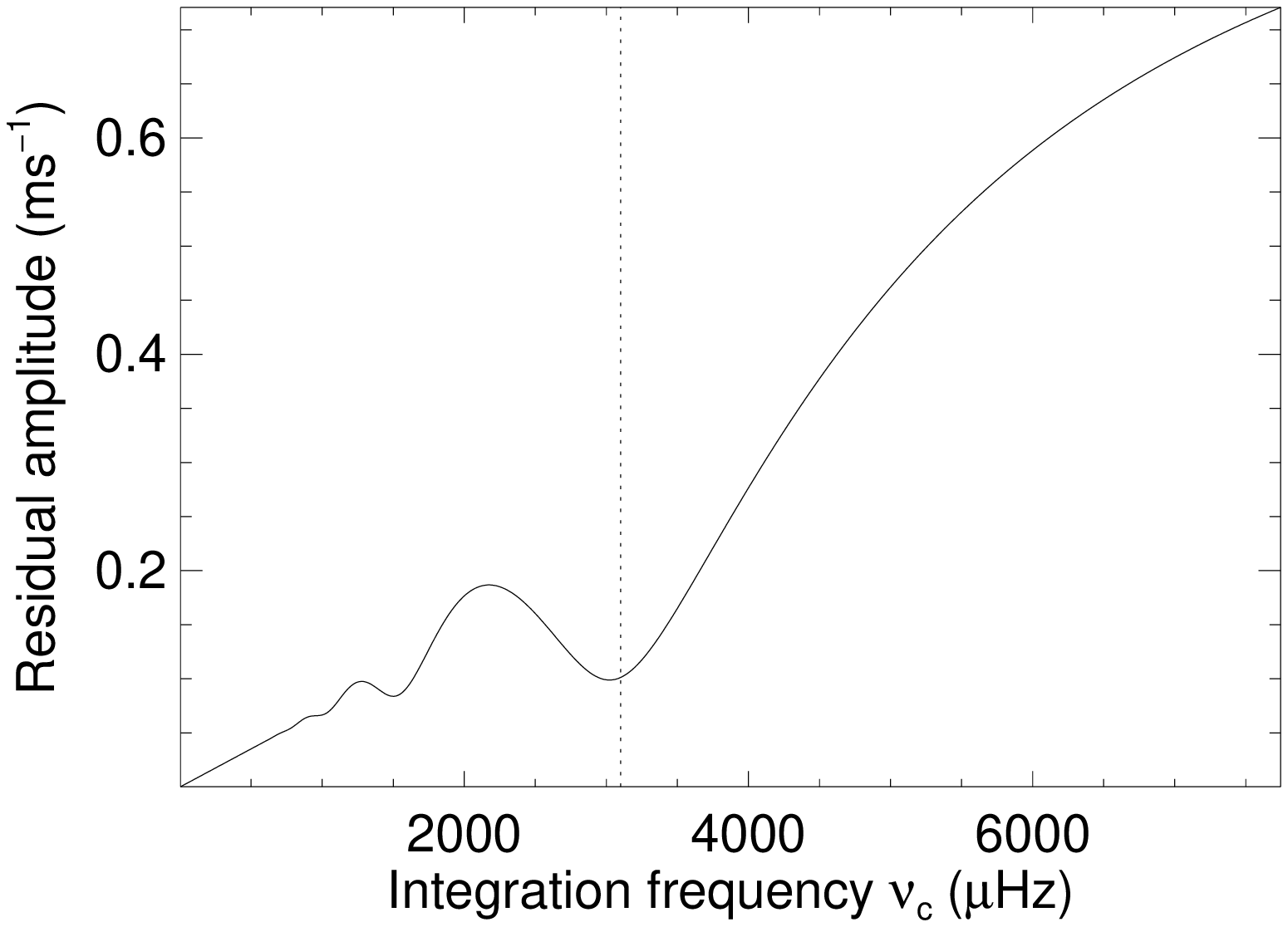}
 \plottwo{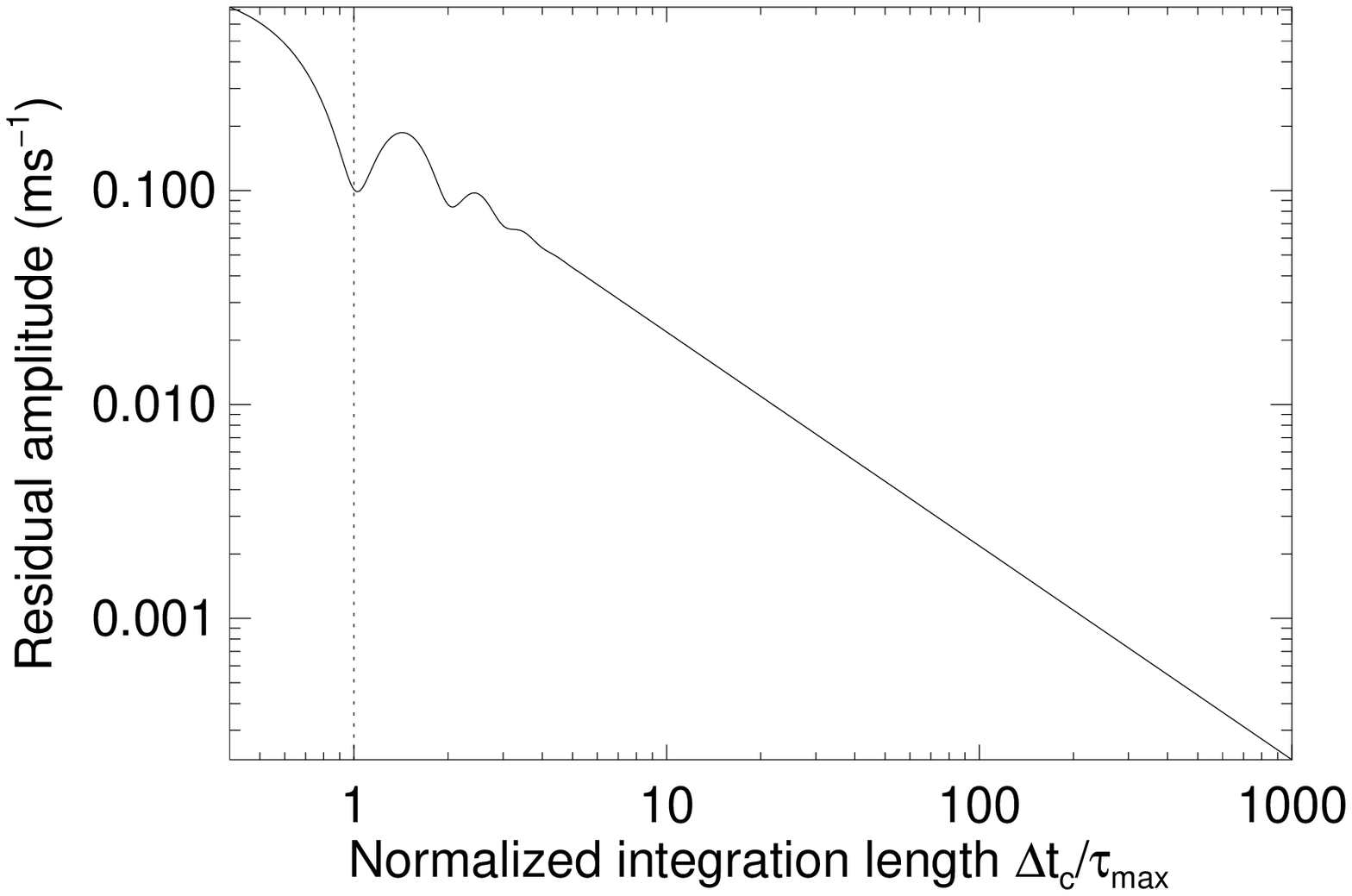}{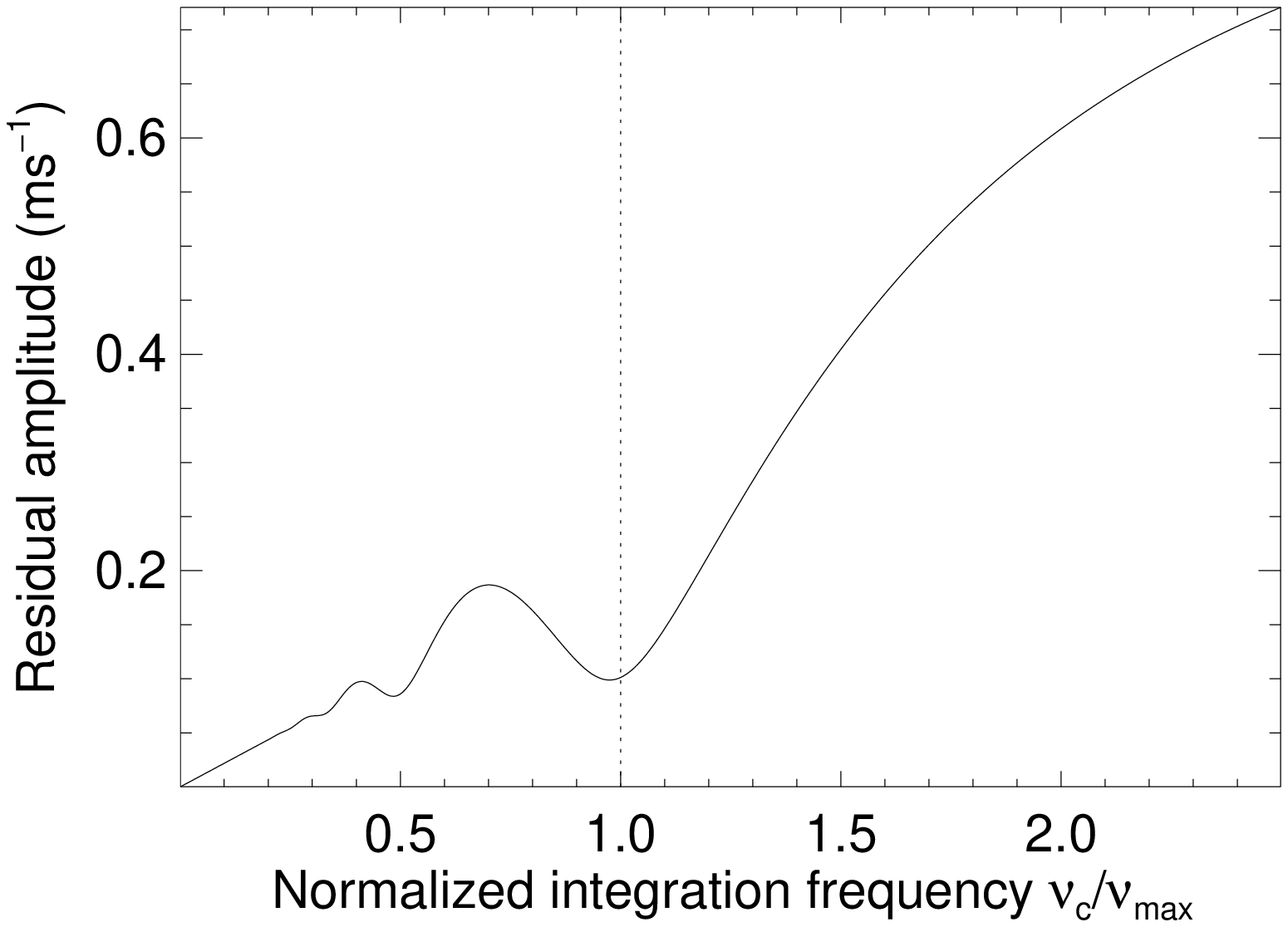}

 \caption{\small Top left-hand panel: residual mode
   amplitude versus integration duration, i.e. the exposure time. The vertical dotted line marks the duration corresponding to $\tau_{\rm max} = 1/\nu_{\rm
     max}$. Top-right hand panel: same as the left-hand, but expressed in the frequency domain. Bottom panels: frequency axes normalized by $\tau_{\rm max}$ and $\nu_{\rm max}$.}

 \label{fig:solar2}
\end{figure}

We may calculate the p-mode signal amplitude that would remain for a given exposure length by multiplying, in frequency, the mode amplitude $\sqrt{P_n}(\nu)$ by the transfer function given by the exposure duration (i.e., the transfer functions like those shown in Fig.~\ref{fig:filt1}). The integral in frequency of this product gives the total remaining or \textsl{residual} mode amplitude. Fig.~\ref{fig:solar2} shows the results for exposures of different duration $\Delta t_{\rm c}$ as applied to the solar spectrum in Fig.~\ref{fig:solar1}. The top left-hand panel shows the residual signal amplitude, in $\rm m\,s^{-1}$, as a function of $\Delta t_{\rm c}$. The vertical dotted line marks the duration corresponding to $\tau_{\rm max} = 1/\nu_{\rm max}$, i.e. corresponding to the peak of the p-mode envelope. The top-right hand panel instead uses the equivalent frequency $\nu_{\rm c} = 1/\Delta t_{\rm c}$ as the independent variable (with $\nu_{\rm max}$ marked by the vertical dotted line). In the bottom panels the frequency axes have been normalized by, respectively, $\tau_{\rm max}$ and $\nu_{\rm max}$.


Whilst these plots show the expected general downward (upward) trend in the residual amplitudes as a function of increasing (decreasing) exposure length (frequency), it is apparent that the residual amplitudes do not fall monotonically at shorter durations (higher equivalent frequencies), where there is pronounced modulation of the response. This can be understood by considering the transfer functions shown in Fig.~\ref{fig:filt1}.

\begin{figure}
 \plotone{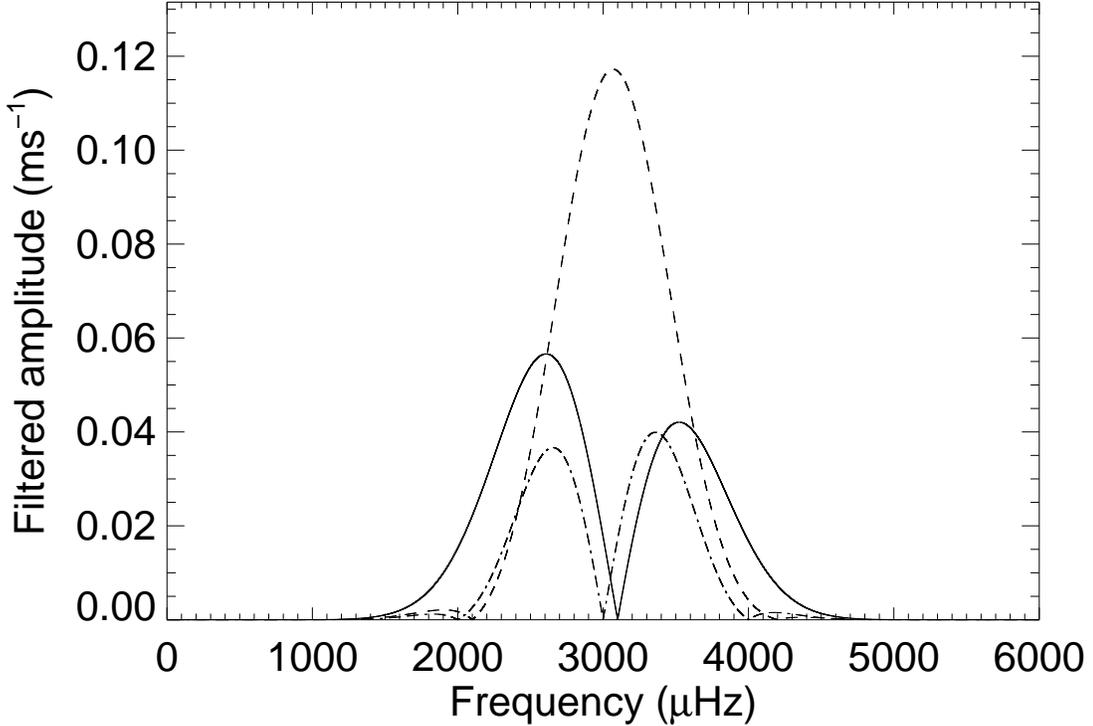}

 \caption{\small Residual amplitudes given by multiplying the frequency responses due to
   different $\Delta t_{\rm c}$ by the mode amplitude due to the
   model solar oscillation spectrum. The solid line shows the resulting residual amplitude 
   for $\Delta t_{\rm c} = \tau_{\rm max} = 5.4\,\rm min$
   (which corresponds to $\nu_{\rm c} = \nu_{\rm max}$). The dashed
   and dot-dashed lines show the responses for $\Delta t_{\rm c}$ of 7.9\,min and 16.7\,min.}

 \label{fig:solar3}
\end{figure}

Fig.~\ref{fig:solar3} shows the result of multiplying each of those filter responses by the mode amplitude in frequency due to the model oscillation spectrum; the solid line shows the residual amplitude when the integration time is equal to $\tau_{\rm max}$, which for our solar mode means $\Delta t_{\rm c} = 5.4\,\rm min$. When $\Delta t_{\rm c} = \tau_{\rm max}$, and therefore $\nu_{\rm c} \equiv \nu_{\rm max}$, the first minimum of the transfer function sits at $\nu_{\rm max}$. 


If the exposure is lengthened one might naively have expected the residual amplitude to drop. However, when $\Delta t_{\rm c} = 7.9\,\rm min$, the integral of the residual signal increases because the first sideband of the sinc-function response is then centered on $\nu_{\rm max}$, as shown in Fig.~\ref{fig:filt1}; this results in a higher fraction of the mode signal being passed. If the duration is lengthened again, say to 16.7\,min, the first sideband shifts to much lower frequencies where the amplitude of the mode signal is much weaker and so the total remaining amplitude drops significantly.


These results indicate that when $\Delta t_{\rm c} = \tau_{\rm max} = 5.4\,\rm min$ the predicted residual amplitude due to the oscillations falls close but not quite to $0.1\,\rm m\,s^{-1}$. Doubling the duration of the exposures has only a modest impact, for the reasons explained above, but does formally reduce the residual amplitude just below the $0.1\,\rm m\,s^{-1}$ level; only when the exposure duration is lengthened again does the predicted residual amplitude once more begin to drop significantly (and then in a monotonic fashion). 

It is important to stress that the curves in Fig.~\ref{fig:solar2} show the expected underlying, noise-free residual amplitude trends. As we shall go on to discuss in the next section, the intrinsic stochastic variability of the solar-like oscillations has the effect of blurring or smearing out the modulation or wiggles in the predicted amplitude trends when only short amounts of data are collected each night, as is usually the case for observations geared to detecting exoplanets.


\section{Results for cool main-sequence, sub-giant and low-luminosity red giants}
\label{sec:other}

The previous section showed results for the Sun (or a solar twin). Making predictions for other stars turns out to be fairly straightforward, because the observed characteristics of the oscillation spectra can be described to reasonable approximation by scaling relations expressed in terms of fundamental stellar properties, involving various combinations of mass, radius, effective temperature, surface gravity and luminosity ($M$, $R$, $T_{\rm eff}$, $g$ and $L$).

As a first cut, given the results from Section~\ref{sec:sun}, one might consider using exposures of duration $\Delta t_{\rm c} = 1/\nu_{\rm max}$ for other stars. It has been shown that $\nu_{\rm max}$ scales to good approximation (e.g., see \citealt{brown91, kjeldsen95, chaplin13}) as
 \begin{equation}
 \nu_{\rm max} \propto MR^{-2} T_{\rm eff}^{-1/2} \propto g\,T_{\rm eff}^{-1/2}.
 \label{eq:numax}
 \end{equation}
However, the maximum amplitudes shown by radial modes in Doppler velocity scale to first-order like (e.g., see \citealt{kjeldsen95, basu17})
\begin{equation}
 A_{\rm max} \propto L/M \propto R^2M^{-1} T_{\rm eff}^4 \propto g^{-1}\,T_{\rm eff}^4 \propto \nu_{\rm max}^{-1} T_{\rm eff}^{7/2}.
 \label{eq:amax}
 \end{equation}
As such, the significant dependence of the amplitudes on stellar properties must also be taken into account. 


To Equation~\ref{eq:amax} we also apply the multiplicative correction of \citet{chaplin11a}, which captures the fact that amplitudes of solar-like oscillations in hotter, F-type stars are suppressed relative to the simple scaling. We note however that our predictions do not reflect that oscillation amplitudes may be suppressed in more active stars \citep{chaplin11b}, and as such may be regarded as upper-limit amplitudes. We return later in this section to discuss the potential impact of stellar cycle variability.

Finally, one other important factor to take account of is that the envelope width of the p-mode oscillation spectrum scales like (see \citealt{mosser12}; Lund et al., in prep):
 \begin{equation} 
 \mbox{$c_{\rm env} \propto$} \left\{
 \begin{array}{lll} 
 \mbox{$\nu_{\rm max}^{0.88}$}&
 \mbox{~~~$T_{\rm eff} < 5777\,\rm K$}\\
 \mbox{$\nu_{\rm max}^{0.88} \left[1+(T_{\rm eff}-5777)/1667\right]$}&
 \mbox{~~~$T_{\rm eff} \ge 5777\,\rm K$.} 
 \end{array} \right.
 \label{eq:cenv} 
 \end{equation}\\

We have adopted solar values of $\nu_{\rm max,\odot} = 3100\,\rm \mu Hz$, $A_{\rm max,\odot} = 0.19\,\rm m\,s^{-1}$, $T_{\rm eff,\odot} = 5777\,\rm K$ and $c_{\rm env,\odot} = 331\,\rm \mu Hz$ to calibrate the relations above.

Equation~\ref{eq:amax} implies that the more evolved (or the more luminous on the main sequence) the star, the larger are its p-mode amplitudes. Whilst $\nu_{\rm max}$ (Equation~\ref{eq:numax}) and hence $c_{\rm env}$ (Equation~\ref{eq:cenv}) decrease as a star evolves the impact of the resulting narrower oscillation envelope is more than offset by the increased mode amplitudes, and the total power in the oscillation envelope increases significantly. 



Here, we ask what is the minimum exposure duration $\Delta t_{\rm c}$ required to reach a residual amplitude in other stars of (i) $0.1\,\rm m\,s^{-1}$; and (ii) the reflex amplitude, $K$, given by an Earth-analogue? We define an Earth analogue as a terrestrial Earth-mass planet (with an Earth-like albedo) that receives the same incident flux from its host star as the Earth receives from the Sun. The Earth-analogue amplitude follows from the relation
 \begin{equation}
 K = K_{\odot}\left(\frac{M}{M_{\odot}}\right)^{-1/2}
 \left(\frac{L}{L_{\odot}}\right)^{-1/4},
 \label{eq:kamp}
 \end{equation}
where $K_{\odot} = 0.09\,\rm m\,s^{-1}$.

\begin{figure*}
 \plottwo{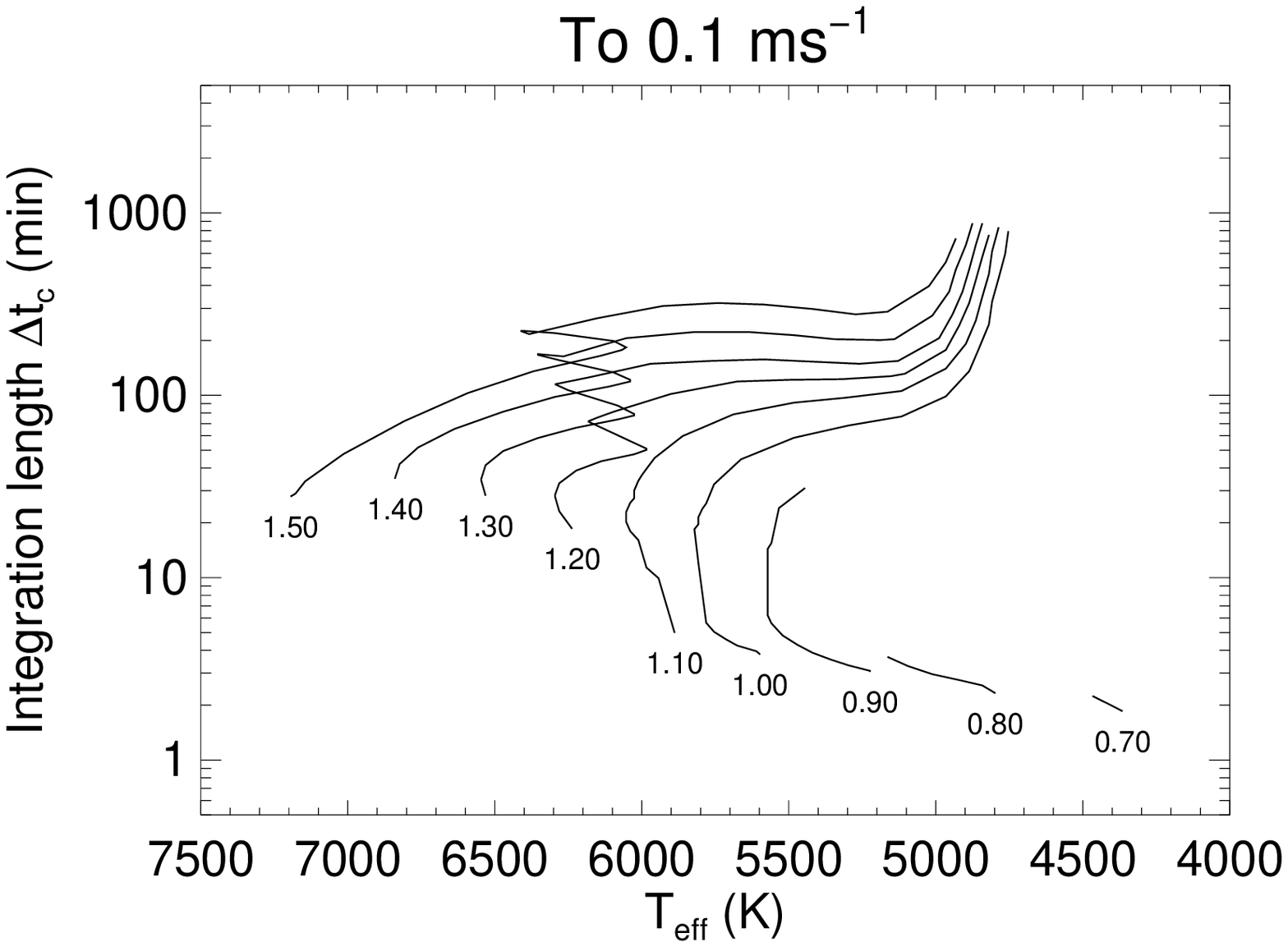}{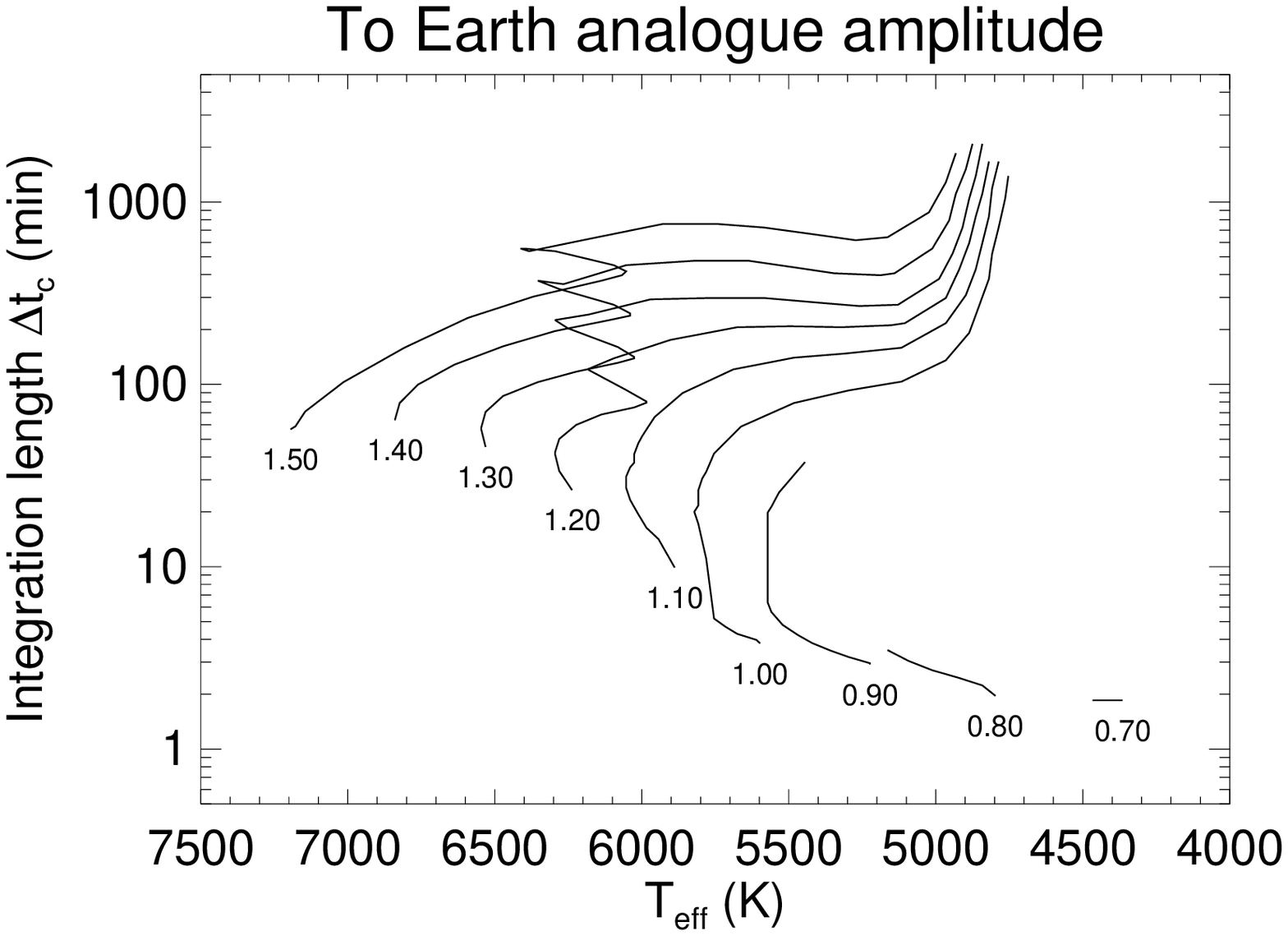}
 \epsscale{0.5}
 \plotone{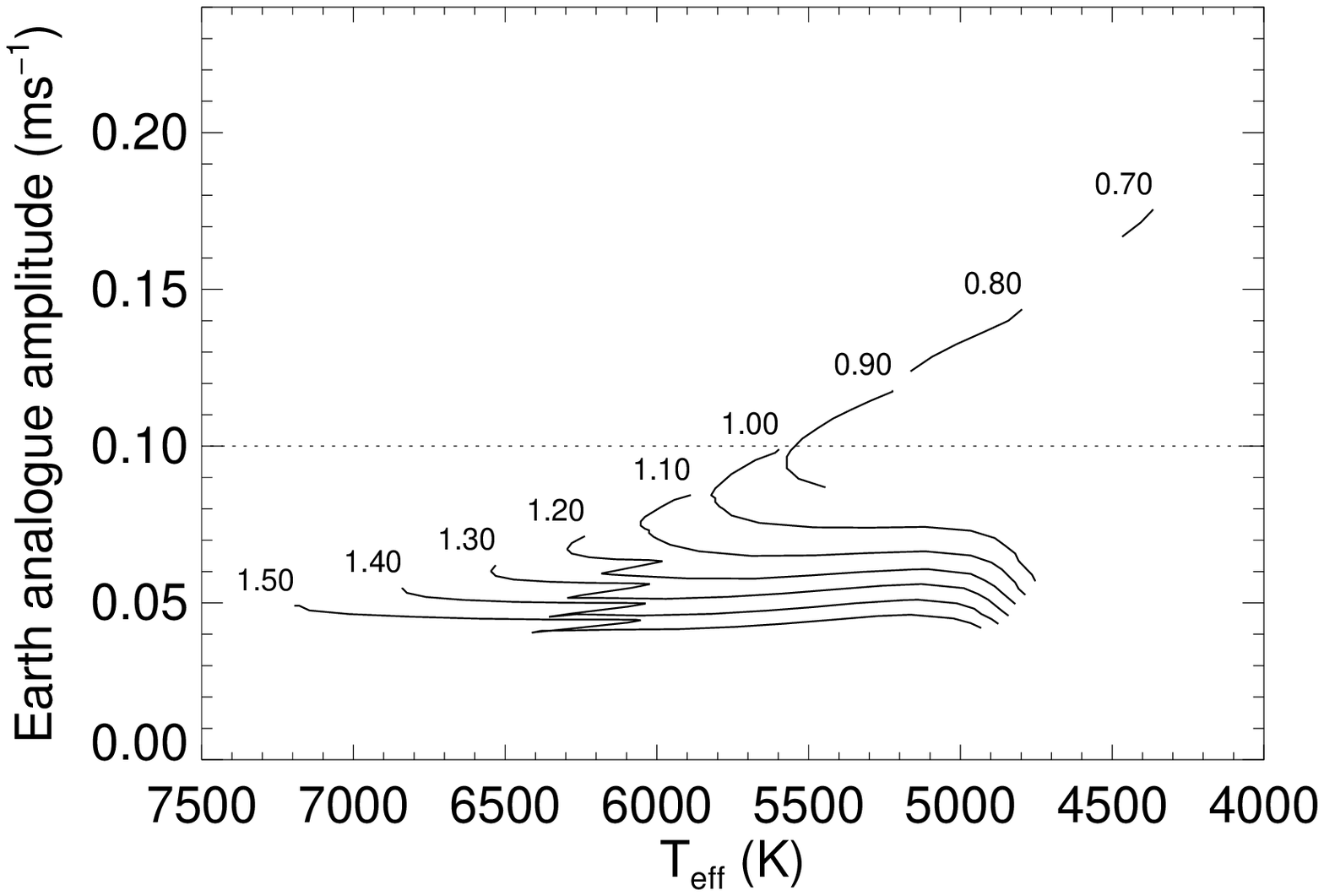}
 \caption{\small Top left-hand panel: Exposure duration $\Delta t_{\rm c}$ needed to
   give a residual amplitude of $0.1\,\rm m\,s^{-1}$, for stellar
   evolutionary tracks of models having masses ranging from $M =
   0.7\,\rm M_{\odot}$ to $M = 1.5\,\rm M_{\odot}$. Top right-hand panel: Exposure   
   duration needed to give a residual amplitude $K$ corresponding to the 
   amplitude given by an Earth analogue. Bottom panel: Values of $K$ for the models along 
   each track. The horizontal dotted line marks the $0.1\,\rm m\,s^{-1}$ threshold.}

 \label{fig:other1}
\end{figure*}

The required exposure durations are plotted in the top two panels of Fig.~\ref{fig:other1}, for stellar evolutionary tracks of models of solar composition (Padova models; see \citealt{marigo08}) from the main sequence through to the base of the red-giant branch having masses ranging from $M = 0.7 - 1.5\,\rm M_{\odot}$. The lower panel shows the Earth-analogue amplitude for models on each of the tracks. 

Both exposure duration plots mimic the classic Hertzsprung-Russell diagram. As expected, longer exposures are needed in sub-giants and low-luminosity red giants. Most striking is that along the main sequence the required durations differ by two orders of magnitude. The Earth-analogue amplitudes $K$ for the hotter (more massive) main-sequence stars are lower than the $0.1\,\rm m\,s^{-1}$ threshold; consequently, a longer exposure is needed to reach the required $K$ thresholds, since more of the lower-frequency parts of the respective oscillation spectra must then be suppressed. The opposite is true for the lower-mass stars.

Note that for the lowest-mass, intrinsically faintest stars in this grid, the shot-noise associated with any observations would likely dominate the signal due to the oscillations, owing to the need to have long exposure times to get sufficient SNR (i.e. longer than the optimal exposure times advocated here), and also because of the very low oscillation amplitudes expected for these stars. As such, we do not bother showing results for stars with $M < 0.7\,\rm M_{\odot}$.

%
%
%


\begin{deluxetable}{llllllcccc}
\label{tab:bright}
\tablecaption{Optimized exposure durations to minimize p-mode amplitudes for a selection of bright stars.}
%
%
%
\tablewidth{0pt}
\tablehead{
 \colhead{Star}&
 \colhead{$M$}& 
 \colhead{$R$}&
 \colhead{$T_{\rm eff}$}&
 \colhead{${\rm log}\,g$}&
 \colhead{$L$}&
 \colhead{Spectral}&
 \colhead{$K$}&
 \colhead{$\Delta t_{\rm c}^{0.1}$}&
 \colhead{$\Delta t_{\rm c}^{K}$}\\
 \colhead{}&
 \colhead{($\rm M_{\odot}$)}&
 \colhead{($\rm R_{\odot}$)}&
 \colhead{(K)}&
 \colhead{(dex)}&
 \colhead{($\rm L_{\odot}$)}&
 \colhead{type}&
 \colhead{($\rm cm\,s^{-1}$)}&
 \colhead{(min)}&
 \colhead{(min)}}
 \startdata
  $\tau$\,Ceti&  0.79&  0.85& 5290&  4.48&  0.51&       G8V& 12.0& $   3.9^{+   4.9}_{-   1.0}$& $   3.7^{+   0.7}_{-   1.3}$\\
$\alpha$\,Cen B& 0.93&  0.91& 5145&  4.49&  0.52&       K1V& 11.0& $   3.6^{+   0.7}_{-   1.1}$& $   3.5^{+   0.7}_{-   1.2}$\\
     70 Oph\,A&  0.89&  0.91& 5300&  4.47&  0.59&       K0V& 10.9& $   4.0^{+   5.1}_{-   1.1}$& $   3.9^{+   0.8}_{-   1.1}$\\
    Sun&  1.00&  1.00& 5777&  4.44&  1.00&       G2V&  9.0& $   9.7^{+   5.9}_{-   5.6}$& $  10.1^{+   7.4}_{-   5.8}$\\
 $\delta$\,Pav&  1.07&  1.20& 5550&  4.31&  1.23&      G8IV&  8.3& $    13^{+    10}_{-   7.8}$& $    16^{+    12}_{-    10}$\\
$\alpha$\,Cen A& 1.11&  1.24& 5745&  4.30&  1.50&       G2V&  7.7& $    14^{+    13}_{-   8.4}$& $    22^{+    13}_{-    10}$\\
  $\iota$\,Hor&  1.23&  1.16& 6080&  4.40&  1.65&       F8V&  7.2& $    13^{+    10}_{-   8.6}$& $    21^{+    11}_{-    10}$\\
    $\mu$\,Ara&  1.21&  1.39& 5665&  4.23&  1.79&    G3IV-V&  7.1& $    21^{+    14}_{-    14}$& $    30^{+    18}_{-    15}$\\
  $\beta$\,Vir&  1.42&  1.69& 6050&  4.13&  3.44&       F9V&  5.5& $    43^{+    23}_{-    21}$& $    80^{+    42}_{-    34}$\\
  $\beta$\,Hyi&  1.08&  1.89& 5790&  3.92&  3.60&  G2IV-G0V&  6.3& $    88^{+    54}_{-    40}$& $   141^{+   119}_{-    58}$\\
 $\alpha$\,CMi&  1.46&  2.13& 6485&  3.95&  7.20&    F5IV-V&  4.5& $   102^{+    64}_{-    45}$& $   203^{+   247}_{-    62}$\\
\enddata
\end{deluxetable}

%
%
%


Table~1 shows, for a selection of bright stars, the minimum exposure durations (in minutes) needed to reach a threshold amplitude of $0.1\,\rm m\,s^{-1}$ ($\Delta t_{\rm c}^{0.1}$) and the corresponding Earth-analogue amplitude, $K$ ($\Delta t_{\rm c}^K$). Note the fundamental properties come from \citet{bruntt10}, and the spectral classifications were taken from SIMBAD\footnote{http://simbad.u-strasbg.fr}. We again see a familiar pattern, that of longer durations being needed for more evolved or more luminous stars.

The relative sizes of the two timescales in Table~1 differ from star to star. The behavior depends on where the threshold amplitude lies with respect to the residual amplitude curves for the star. The curves for the Sun were shown in Fig.~\ref{fig:solar2}.  If the threshold amplitude is in the range where we see pronounced modulation of the residual amplitude curves at short exposure durations, the required durations do not vary monotonically (as noted previously in Section~\ref{sec:sun}). For the several centimeter-per-second threshold levels considered in the paper, this effect is most important for stars with masses around 0.9 to $1.1\,M_{\odot}$. It explains the small wiggles seen in the corresponding tracks shown in the top two panels of Fig.~\ref{fig:other1}. When the threshold amplitude lies in the monotonically varying part of the residual amplitude curves at longer durations, the behavior is more straightforward.



The exposure duration uncertainties given in Table~1 take several factors into account. First, there are estimated uncertainties in the fundamental stellar properties, which are inputs to the predictions. For these bright stars this contributes a median uncertainty in the exposure durations of around only 5\,\%. Next, there are uncertainties in the scaling relations, e.g., equivalent 1-$\sigma$ uncertainties of about $15\,\%$ in $A_{\rm max}$ and $c_{\rm env}$ (e.g., see \citealt{chaplin11a, campante14, lund17}). Some of this uncertainty is undoubtedly statistical, owing to the finite precision of the \emph{Kepler} data used to derive the relations, and therefore not intrinsic to the scalings themselves.

Power in the solar-like oscillations is also affected by changing levels of near-surface magnetic activity. As the Sun moves from the minimum to maximum phase of its 11-year cycle, the \textsc{rms} amplitudes of the most prominent low-$l$  p modes are reduced by about 10\,\% \citep{elsworth93,chaplin00,komm00}. Similar levels of change have now been found in other Sun-like stars observed by \emph{Kepler} (e.g., see \citealt{garcia10,kiefer17,salabert18}). This activity-related variability should already be captured by the fits used to constrain the scaling relations, because the \emph{Kepler} data have sampled stars at different phases of their cycles. Very active stars are unlikely to play a significant role: as noted previously, the amplitudes of their oscillations are heavily suppressed \citep{chaplin11b}, and so they are selected against in any sample of targets used to constrain the relations. Very active stars are anyway challenging for exoplanet searches. We note also that there is no evidence in the literature from available seismic data on Sun-like stars for any significant dependence of mode amplitudes on metallicity; however, as per the impact of activity, any spread due to this would be captured by the uncertainties on the existing scaling relations.

However, the most important uncertainty in practical terms is that arising from the stochastic excitation and damping of the solar-like oscillations. Mode amplitudes show significant variability when measured on timescales that are shorter than the mode lifetimes \citep{jcd01}. Those conditions are satisfied here, i.e., relevant nightly observations are of few-hour or shorter durations, versus typical lifetimes of several days or more for the oscillations. It is important to stress that this variability is intrinsic to the modes, and will be present even if the shot-noise is extremely low. To give a feel for the extent of this variability, consider the example of a Sun-like oscillation spectrum. The measured total \textsc{rms} oscillation signal will show intrinsic scatter at the $\simeq 50\,\%$ level for 5-min integrations, and at the $\simeq 35\,\%$ level for 15 to 20-min exposure durations. The data in Table~1 take this intrinsic variability into account, via realistic Monte Carlo simulations of the predicted oscillation spectra, and it has the effect of smearing out or blurring the sinc-induced modulation (or wiggles) shown by the residual mode amplitude curves in Fig.~\ref{fig:solar2}. Note the exposure durations listed in the table take this blurring into account (they are the median exposure durations from the Monte-Carlo simulations).

The combined effect of all of the above sources (assumed to be uncorrelated) is to give median fractional positive and negative uncertainties (68\,\% equivalent) in the exposure durations of around 55 to 65\,\% (positive) and 40 to 50\,\% (negative), respectively.

\begin{figure*}
 \plotone{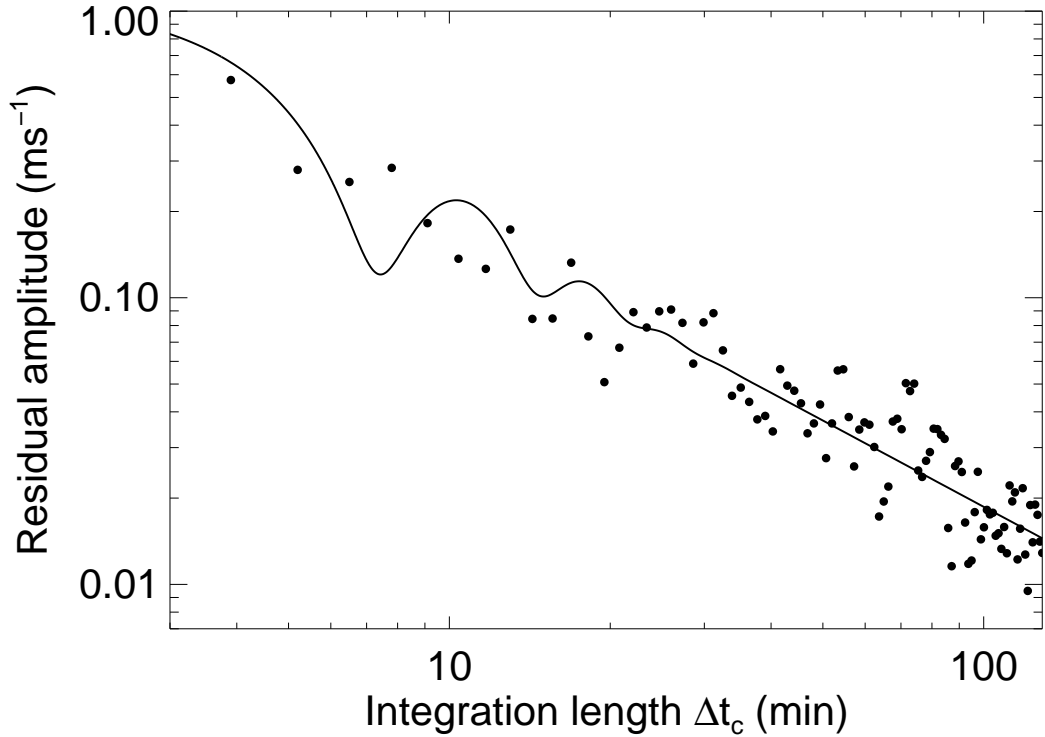}

 \caption{\small Residual mode amplitude versus integration duration, i.e. the exposure time, showing real results from observations of $\alpha$\,Cen A (filled symbols) and the model prediction (solid line).}

 \label{fig:acen1}
\end{figure*}

Fig.~\ref{fig:acen1} shows the residual amplitudes given by 8\,hours of high-cadence Doppler data collected by \citet{butler04} on the bright Sun-like star $\alpha$\,Cen~A, using UVES on the VLT. Note we first applied a 1-hour high-pass filter to the data to remove longer-term trends due to granulation, activity and instrumental drifts, and applied a small correction for the shot-noise. The filled symbols then show the residual signal amplitudes given by re-binning to integrations $\Delta t_{\rm c}$ of increasing duration. The solid line is our model prediction, based on the data in Table~1. The real data follow very nicely the underlying trend of the model, but are scattered around it for the reasons described above, i.e., when we have just a few hours of data, the intrinsic stochastic variability blurs out the wiggles in the underlying curve.


\section{How to use these results}
\label{sec:how}

To enable the community to use our results we have made publicly available a \texttt{Python} code that computes the exposure durations $\Delta t_{\rm c}^{0.1}$ and $\Delta t_{\rm c}^K$ given basic stellar observables as inputs. Code and documentation are available at \texttt{https://github.com/grd349/ChaplinFilter} and may be installed using \texttt{PyPi} \footnote{\texttt{pip install chaplinfilter}}.  

The \texttt{ChaplinFilter} code replicates the map from stellar input parameters to the output exposure durations using a supervised learning technique. In order to achieve a good precision and accuracy on the outputs we selected three inputs to train on: luminosity, $L$; surface gravity, ${\rm log}\,g$; and effective temperature, $T_{\rm eff}$. These are data that should be readily available for targets of interest, in particular given the recent Gaia DR2 release. We used a random forest regression to learn the relation between $\{{\rm log}\,g, T_{\rm eff}, L\}$ and $\Delta t_{\rm c}^{0.1}$ or $\Delta t_{\rm c}^K$, using the data in the top two panels of Figure~\ref{fig:other1} as the training input.  The trained algorithm consistently achieves an out-of-bag score $R^2$ of better than 0.98, where $R^2$ is defined as
\begin{equation}
R^{2} = 1 - \frac{\sum_{i} (y_{\rm true, i} - y_{\rm pred, i})^2}{\sum_{i} (y_{\rm true, i} - \left<y_{\rm true}\right>)^2}.
\end{equation}
The input-output maps for each exposure duration are accurate typically to a level of a few per cent, which given the typical uncertainty expected on the inputs is deemed to be acceptable. Most of the information on the output timescales is provided by $L$ (most important) and ${\rm log}\,g$, with $T_{\rm eff}$ providing only a small amount of additional constraint. This is not surprising, given that the first two observables already capture information on temperature, radius and mass.

The \texttt{OscFilter} code provides a callable and deterministic function that calculates the outputs given the required inputs. We then recommend considering uncertainties of the magnitude listed in Section~\ref{sec:other}.

\section{Conclusions}
\label{sec:conc}

Cool stars show solar-like (p-mode) oscillations, which are a crucial source of data for providing extremely precise and accurate stellar properties, information of fundamental importance for planet-hosting stars. The oscillations, however, also present an important source of astrophysical noise in searches for low-mass or Earth-analogue planets. In this paper we have explored how fine-tuning the exposure durations in radial velocity searches can effectively average out the p-mode oscillations. We make recommendations for the appropriate exposure lengths to use, depending on the stellar properties of the target, and provide an easy-to-use \texttt{Python} code that computes guideline exposure durations given basic stellar observables as inputs to reduce the residual oscillation amplitudes to $\simeq 0.1\,\rm m\,s^{-1}$ and the equivalent Earth-analogue amplitude for the star. Owing to the intrinsic stochastic variability of the oscillations, we recommend in practice choosing short exposure durations at the telescope and then averaging over those exposures later, as guided by our predictions.

For the Sun, the relevant integration times are in the range 10\,min. For cooler, low-mass stars very low residual amplitudes can be obtained with even shorter exposures (as brief as 4\,minutes for K-type dwarfs). However, to reach similar residuals amplitudes for hotter stars and sub-giants demands that exposure times be up a few tens of minutes or even longer (over 100\,minutes). 

The most important message of our paper is that exposure times must be carefully tailored, depending on the stellar properties. This matters not only in terms of optimizing the precision, but also because we could be wasting precious telescope time by collecting data using non-optimal exposures. This is particularly relevant for transit observations, e.g., for extracting subtle signatures due to the Rossiter-McLaughlin effect or exoplanet atmospheres.

%
%
\acknowledgments

W.J.C., G.R.D. and C.A.W. acknowledge support from the UK Science and Technology Facilities Council (STFC). W.J.C., G.R.D. and W.H.B. also acknowledge support from the UK Space Agency.  H.M.C. acknowledges the financial support of the National Centre for Competence in Research (NCCR) “PlanetS” supported by the Swiss National Science Foundation (SNSF). Funding for the Stellar Astrophysics Centre is provided by The Danish National Research Foundation (Grant agreement no.: DNRF106). The authors thank Tim Bedding for providing the Doppler velocity data on $\alpha$\,Cen A, and also the anonymous referee for helpful comments and suggestions.


\end{document}